\pdfoutput=1
\documentclass[conference,a4paper]{IEEEtran}
\IEEEoverridecommandlockouts
\usepackage{cite}
\usepackage{siunitx}
\usepackage{amsmath,amssymb,amsfonts}
\usepackage{algorithmic}
\usepackage{graphicx}
\usepackage{float}
\usepackage{subcaption}
\usepackage{textcomp}
\usepackage{multicol}
\usepackage{xcolor}
\usepackage{lipsum}
\usepackage{tikz}
\usetikzlibrary{bayesnet}
\def\BibTeX{{\rm B\kern-.05em{\sc i\kern-.025em b}\kern-.08em
    T\kern-.1667em\lower.7ex\hbox{E}\kern-.125emX}}

\makeatletter
 \let\old@ps@headings\ps@headings
 \let\old@ps@IEEEtitlepagestyle\ps@IEEEtitlepagestyle
 \def\confheader#1{%
 \def\ps@headings{%
 \old@ps@headings%
 \def\@oddhead{\strut\hfill#1\hfill\strut}%
 \def\@evenhead{\strut\hfill#1\hfill\strut}%
 }%
 \def\ps@IEEEtitlepagestyle{%
 \old@ps@IEEEtitlepagestyle%
 \def\@oddhead{\strut\hfill#1\hfill\strut}%
 \def\@evenhead{\strut\hfill#1\hfill\strut}%
 }%
 \ps@headings%
 }
\makeatother

\confheader{
2021 International Conference on Indoor Positioning and Indoor Navigation (IPIN), 29 Nov. -- 2 Dec. 2021, Lloret de Mar, Spain}

\IEEEoverridecommandlockouts
\IEEEpubid{\makebox[\columnwidth]
{\hfill 978-1-6654-0402-0/21/\$31.00~\copyright~2021 IEEE}
\hspace{\columnsep}\makebox[\columnwidth]{ }}	

\begin{document}

\title{Measuring Uncertainty in Signal Fingerprinting 
\\ with Gaussian Processes Going Deep
}

\author{
    \IEEEauthorblockN{Ran Guan\IEEEauthorrefmark{1}, Andi Zhang\IEEEauthorrefmark{2}\IEEEauthorrefmark{3}, Mengchao Li\IEEEauthorrefmark{1} and Yongliang Wang\IEEEauthorrefmark{1}}
    \IEEEauthorblockA{\IEEEauthorrefmark{1}
    \textit{Riemann Laboratory} \\
    \IEEEauthorrefmark{3}
    \textit{Noah's Ark Laboratory} \\
    \textit{2012 Laboratories, Huawei}
    \\\{guanran, limengchao, wangyongliang775\}@huawei.com}
    \IEEEauthorblockA{\IEEEauthorrefmark{2}
    \textit{Department of Computer Science and Technology} \\
    \textit{University of Cambridge}
    \\\{az381\}@cam.ac.uk}
}


\maketitle

\begin{abstract}
In indoor positioning, signal fluctuation is highly location-dependent. However, signal uncertainty is one critical yet commonly overlooked dimension of the radio signal to be fingerprinted. This paper reviews the commonly used Gaussian Processes (GP) for probabilistic positioning and points out the pitfall of using GP to model signal fingerprint uncertainty. This paper also proposes Deep Gaussian Processes (DGP) as a more informative alternative to address the issue. How DGP better measures uncertainty in signal fingerprinting is evaluated via simulated and realistically collected datasets. 
\end{abstract}

\begin{IEEEkeywords}
Indoor Positioning, Signal Fingerprinting, Gaussian Processes, Deep Gaussian Processes, Uncertainty, Heteroscedasticity
\end{IEEEkeywords}

\section{Introduction}
Signal Fingerprinting as an infrastructure-free approach for indoor positioning has been widely investigated over the decades. Typical fingerprinting consists of two major steps, the offline surveying stage and the online positioning stage. In the offline surveying stage, groups of radio signal observations, typically WiFi RSSI (Received Signal Strength Index), are annotated with location labels either by dedicated surveyors or via crowdsourcing\cite{guan2017towards} to form fingerprints. In the online positioning stage, a freshly collected fingerprint needs to be matched to those collected to be positioned. For a detailed review of signal fingerprinting, we refer our readers to works such as \cite{he2016wi,xia2017indoor}. Major fingerprinting approaches can be broadly categorised as deterministic or probabilistic. Deterministic approaches such as K-Nearest-Neighbours (KNN) are straightforward to implement. However, they typically lack the capability to predict fingerprints outside surveyed area and do not model the uncertainty of fingerprints. Signal propagated indoors exhibit high degrees of uncertainty and failing to capture the uncertainty accurately potentially means a deterioration in positioning accuracy. Measuring uncertainty also helps to indicate localisability \cite{guan2018signal} or useability of the positioning service. 

By contrast, GP \cite{rasmussen2006gaussian} handles these issues and models signal fingerprints from a probabilistic perspective. The key idea of GP is to learn a spatial-signal relationship from the collected fingerprints through the kernel a.k.a. the covariance function. With the spatial-signal relationship learned or abstracted, GP enables fingerprint prediction at the unexplored area. Due to its probabilistic nature, GP is also widely considered to be able to measure signal uncertainty. However, we found that the standalone GP does not model the uncertainty in fingerprinting context as most would have expected. To illustrate this, we first walk our readers through a generic probabilistic setup of the fingerprint positioning problem and demonstrate the heteroscedasticity of the signal via a contrived dataset in Section \ref{sec:setup}. In Section \ref{sec:gp}, we review GP and show that the standalone GP does not reflect the uncertainty of the signal directly. We introduce DGP \cite{damianou2013deep}, a multi-layer network with GP stacked together as an alternative to better measure the uncertainty from the collected fingerprints in Section \ref{sec:dgp}. We justify the use of DGP using a simulated fingerprint dataset and multiple realistically collected datasets in Section \ref{sec:eval}. A comparison of how well GP and DGP measure the uncertainty is also made in the same section. Section \ref{sec:conclusion} concludes our paper with implications for future indoor positioning system design.

The contribution of this paper is multifold:
\begin{enumerate}
  \item We demonstrate the heteroscedasticity of signal fingerprints via an experiment that can be easily reproduced elsewhere, which, to our best knowledge, has never been formally studied before;
  \item We show, both theoretically and empirically, that the commonly practised GP based fingerprinting fails to model uncertainty correctly due to its homoscedastic assumption;
  \item We introduce and apply DGP to signal fingerprinting to better measure uncertainty; 
  \item We compare DGP and GP on multiple datasets and show how accurate uncertainty measurement helps to improve the accuracy of current fingerprinting based indoor positioning. 
\end{enumerate}

\section{Probabilistic Fingerprinting}\label{sec:setup}

We begin by framing the fingerprinting problem in a probabilistic setting. Assume the area of interest can be organised as a chessboard, with $M$ rows and $N$ columns, making in total $M \times N$ grids. On this chessboard, a smart device can report the RSSI measurements from up to $K$ different WiFi access points (APs), whose locations in the area are unknown. At each grid, the RSSI follows a certain probabilistic distribution $\mathbb{P}$ with a mean $\mu$ and a variance $\sigma^2$. During the offline surveying stage, surveying devices traverse, typically just part of, the area and form a training fingerprint dataset where the RSSI measurements can be considered as sampled results from these distributions. For example, at grid $\mathbf{x} = (m,n)\in\mathbb{R}^2$, the device can hear an RSSI value of $y_{mnk}\in\mathbb{R}$, sampled from the distribution of the $k$-th AP at that grid. In other words, $y_{mnk} \sim \mathbb{P}_{mnk}(\mu_{mnk},\sigma_{mnk}^2)$. Therefore, a surveyed fingerprint is a location tag and signal observation pair. Given a newly measured collection of RSSI observations $Y = (y_1,y_2,\dots,y_K)$ at the online positioning stage, the task is to recover its location tag. Notice that the uncertainty of the signal refers to the variance $\sigma^2$ of the distribution of the corresponding AP and the grid.

\subsection{Assumptions and Positioning}

Most works assume the RSSI distribution $\mathbb{P}$ is Gaussian. This does not hold since RSSI is not uniquely determined by the location but also by temporal factors such as sporadic people movement nearby. For the same reason, distributions like Rayleigh or even Rice cannot fully characterise RSSI, particularly in challenging situations. However, Gaussian $\mathbb{P}$ is assumed for simplicity as it allows easy manipulation of the likelihood of observing $y$ at location $\mathbf{x} = (m,n)$ as:
\begin{equation}\label{eq:likeli}
    p(y|\mathbf{x}) = \frac{1}{\sigma_{mn}\sqrt{2\pi}} 
  \exp\left( -\frac{1}{2}\left(\frac{y-\mu_{mn}}{\sigma_{mn}}\right)^{\!2}\,\right)
\end{equation}

  Subsequently, the posterior probability is given as:
\begin{equation}\label{eq:posterior}
    p(\mathbf{x}|y) = \frac{p(y|\mathbf{x})p(\mathbf{x})}{p(y)} = \frac{p(y|\mathbf{x})p(\mathbf{x})}{\sum_{i=1,j=1}^{M,N} p(y|(i, j))p((i,j))} 
\end{equation}

In Equation \ref{eq:posterior}\footnote{The notation $p(y|(i,j))$ is short for $\mathbb{P}(Y=y|X=(i,j))$.}, the $p(\mathbf{x})$ term is the prior probability of the location. Linking this term to a previously located coordinate allows tracking. Here we only care about one-shot (push-to-fix) positioning and assume a uniform distribution over the chessboard. Additionally, the $p(y)$ term is the marginal likelihood and can be calculated by the Law of total probability, but it can be omitted as it stays fixed with reference to any given $y$. As a result, we have:
\begin{equation}\label{eq:approx_posterior}
    p(\mathbf{x}|y) \propto  p(y|\mathbf{x})
\end{equation}

Another important assumption is that the RSSI distributions from different APs are independent of each other. While this does not hold in cases where virtual APs with different mac addresses actually share the same physical AP, modelling them separately does not harm the positioning result in theory. Therefore the maximum a posteriori (MAP) estimation of the location is:
\begin{equation}\label{eq:map}
    \hat{\mathbf{x}} = \mathop{\arg\max}_{m,n} \left ( \prod_{k=1}^{K} p(y_k|(m,n)) \right )
\end{equation}

Under the above probabilistic framework, the positioning outcome hinges on the construction of the distributions of RSSI over the chessboard (Equation \ref{eq:likeli}). The fingerprint positioning problem can be now solved by learning the $ \mu $ and $ \sigma^2 $ across all grids and APs from the surveyed fingerprints, which is also referred to as the recovery of radio maps. 

\subsection{Heteroscedasticity}
It is worth emphasising that the uncertainty (modelled by $ \sigma^2 $) of the signal contributes to the positioning result, as indicated by Equation \ref{eq:map}. Another way to see this is to consider that the signal is heteroscedastic, that is to say, the variance is spatially varying. Spatially varying properties or signals naturally help to anchor positions.

\begin{figure}[h]
\centerline{\includegraphics[width=0.45\textwidth]{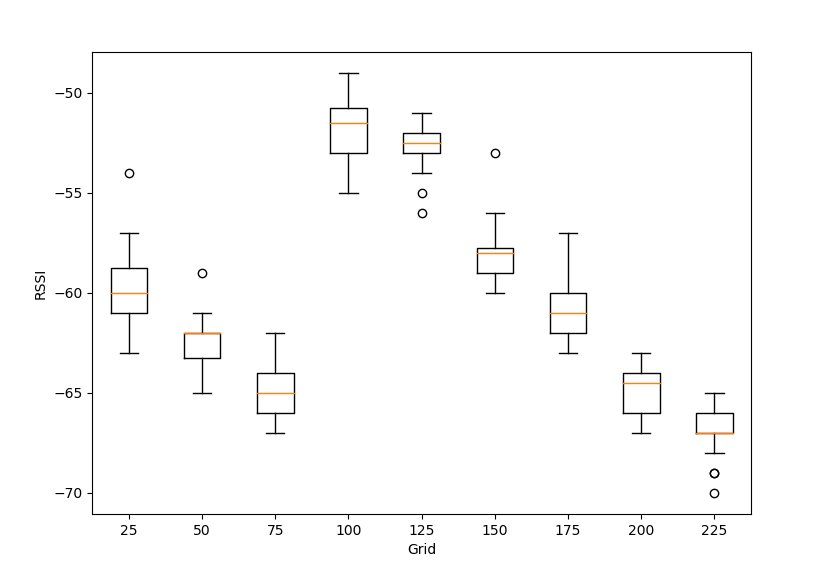}}
\caption{Box plots of RSSI Signals of an AP at 9 Different Grid Positions.}
\label{fig:rssi_boxplot}
\end{figure}

To verify the heteroscedasticity of RSSI measurements, we contrived a simple fingerprint dataset in an average office building with a Huawei Mate 30 smartphone as the surveying device. Fingerprints were collected at 225 different positions during the working hours within one day. Each position contributes 20 fingerprints and is at least 1 meter apart from other positions. Fingerprints collected at the same positions can be considered as measurements sampled from the underlying distributions. Among all 4500 fingerprints, 26 APs were audible consistently. We tested for the heteroscedasticity of these 26 APs using Levene's test\cite{derrick2018tests} and the results indicate heteroscedasticity for all APs. Intuitively, the spatial difference in variance can also be identified visually by box plots such as in Figure \ref{fig:rssi_boxplot}.

We believe this test can be readily repeated in other settings since the uncertainty links greatly to environmental factors. Heteroscedasticity of the wireless signal suggests that the variance is location related thus embeds hints on positioning. Inaccurate modelling of the variance may lead to a loss of accuracy. However, the heteroscedasticity of RSSI is rarely formally discussed by the fingerprinting community. In convention, the variance information is either poorly modelled by deterministic approaches or implicitly modelled as homoscedastic by GP based approaches.

\section{Gaussian Processes}\label{sec:gp}

 GP has been promoted in signal fingerprinting for two major reasons\cite{hahnel2006gaussian,dashti2015rssi,richter2015revisiting,yiu2016locating,zhao2016gaussian,KUMAR20161,yiu2017wireless,sun2018augmentation}. Firstly, it is able to interpolate or predict RSSI at the unexplored area. Secondly, it is considered to be able to model the RSSI uncertainty. Unfortunately, we found that the commonly practised GP only models RSSI variance as a mixture of homogeneous noise and surveyed fingerprint coverage instead of the uncertainty as described in Section \ref{sec:setup}, which would greatly limit the claimed advantages of GP in signal fingerprinting. 

Following the framework described in Section \ref{sec:setup}, GP typically models RSSI observations as samples drawn from a noisy process with a zero-mean Gaussian noise:
\begin{equation}\label{eq:gp_noise}
y = f(\mathbf{x}) + \epsilon
\end{equation}

There are two key ingredients to GP, the mean function and the kernel. The mean function defines the prior knowledge of the expected outputs and is assumed to be zero in most works. The kernel captures the intuition that similar inputs (in this context grid positions) are likely to result in similar outputs (RSSI observations). The squared exponential (SE) kernel is a widely applied option and has the form of:
\begin{equation}\label{eq:gp_kernel}
	k(\mathbf{x}_p,\mathbf{x}_q) = \sigma^2_f\exp(-\frac{1}{2l^2}||\mathbf{x}_p - \mathbf{x}_q||^2)
\end{equation}

Here $l$ is the characteristic length scale that can be thought of as the spatial distance needed to be travelled before a significant change in RSSI can be observed. The $\sigma^2_f$ is (somewhat misleadingly) described as the variance of the signal in many works. Together with the variance $\sigma^2_n$ of the Gaussian noise $\epsilon$, these three terms are referred to as the hyperparameters $\theta = (l, \sigma^2_f , \sigma^2_n)$ and model the covariance between noisy signal observations:
\begin{equation}\label{eq:gp_cov_noise}
	cov(y_p,y_q) = k(\mathbf{x}_p,\mathbf{x}_q)  + \sigma^2_n\delta_{pq}
\end{equation}

where $\delta_{pq}$ is one iff $p=q$ and zero otherwise. Given a set of positions of surveyed fingerprints $\mathbf{X}\in \mathbb{R}^{N\times 2}$ ($N$ is the number of fingerprints), we have the covariance over the corresponding RSSI observations $\mathbf{Y}\in \mathbb{R}^N$ (from one AP) as:
\begin{equation}\label{eq:gp_cov_mat}
	cov(\mathbf{Y}) = K(\mathbf{X}) + \sigma^2_nI
\end{equation}

Here $K(\mathbf{X})$ denotes the covariance matrix of $\mathbf{X}$, whose element $K(\mathbf{X})_{pq}$ is $k(\mathbf{x}_p,\mathbf{x}_q)$. Finally, RSSI at an arbitrary position $\mathbf{x}_*$ can be derived via the key predictive equations of GP:
\begin{equation}\label{eq:gp_post_noise}
	p(y_*|\mathbf{x_*},\mathbf{X},\mathbf{Y}) = \mathcal{N}(y_*;\mu_{X_*},\sigma^2_{X_*} + \sigma^2_n)
\end{equation}
\begin{equation}\label{eq:gp_post_mean}
	\mu_{x_*} = \mathbf{k_*}^T(K(\mathbf{X}) + \sigma_n^2I)^{-1}\mathbf{Y}
\end{equation}
\begin{equation}\label{eq:gp_post_variance}
	\sigma^2_{x_*} = k(\mathbf{x}_*,\mathbf{x}_*) - \mathbf{k_*}^T(K(\mathbf{X}) + \sigma_n^2I)^{-1}\mathbf{k_*} 
\end{equation}

where $\mathbf{k}_*$ denotes the vector of covariances between $\mathbf{x}_*$ and $\mathbf{X}$. The log marginal likelihood of all fingerprints given the hyperparameters is: 
\begin{flalign}\label{eq:gp_loglikeli}
\begin{split}
 \log\, p(\mathbf{Y}|\mathbf{X};\theta) = \\
 & -\frac{1}{2}\mathbf{Y}^T(K(\mathbf{X})+\sigma^2_nI)^{-1}\mathbf{Y} \\&-
 \frac{1}{2}\log|K(\mathbf{X})+\sigma^2_nI|-\frac{N}{2}log2\pi
 \end{split}
\end{flalign}

By optimising Equation \ref{eq:gp_loglikeli} with reference to each hyperparameter, GP based fingerprinting can approximate Equation \ref{eq:likeli} via Equation \ref{eq:gp_post_noise} with global optimal hyperparameters. 

Full details of GP are available elsewhere \cite{rasmussen2006gaussian,hahnel2006gaussian,ferris2007wifi,KUMAR20161}. In particular, we ask our readers to pay attention to the variance of the predictive distribution of the above GP approach. In Equation \ref{eq:gp_post_noise}, the variance of the predictive Gaussian distribution consists of two parts, the spatially variant term $\sigma^2_{x_*}$ (Equation \ref{eq:gp_post_variance}) and the spatially invariant term $\sigma^2_n$, that is fixed by definition (in Equation \ref{eq:gp_noise}, the noise term $\epsilon$ is modelled as independent and identically distributed across all grids) and is learned by optimising Equation \ref{eq:gp_loglikeli}. Notice that the resulted predictive model is still heteroscedastic as a whole since $\sigma^2_{x_*}$ is spatially varying. However, $\sigma^2_{x_*}$ does not depend on $\mathbf{Y}$ (as indicated by Equation \ref{eq:gp_post_variance}). Instead, it only depends on $\mathbf{X}$ (location tags of the surveyed fingerprints). The first term in Equation \ref{eq:gp_post_variance} is in fact a fixed term with the learned value of $\sigma^2_f$, and the second term monotonously increases with the size of $\mathbf{X}$ but saturates before it reaches $\sigma^2_f$, which means $\sigma^2_{x_*}$ does not reflect the signal uncertainty (the spread of RSSI observations at $x_*$) and is a measure of how well nearby area has been surveyed. Hence the ultimate predictive variance ($\sigma^2_{x_*}$ + $\sigma^2_n$) consists of two terms that are not related to the signal uncertainty. 


We attribute the misuse of GP mainly to the misunderstanding that the signal is homoscedastic. On the contrary, RSSI signal is highly heteroscedastic as shown in Section \ref{sec:setup}, yet GP models noise terms as additive i.i.d. with reference to locations (Equation \ref{eq:gp_noise}). GP assumes homoscedasticity but the predictive distribution resulted (Equation \ref{eq:gp_post_noise}) is heteroscedastic, which may have misled many to believe that the variances of the predictive distributions measures spatially-varying uncertainties. In addition to the loss of accuracy at surveyed grids due to inaccurate modelling of signal uncertainty, a subtle yet more worrying consequence is that the positioning result will always be biased towards surveyed areas since $\sigma^2_{x_*}$ is solely determined by the coverage of nearby collected fingerprints. How this affects positioning accuracy in practice will be investigated in Section \ref{sec:eval}.

\begin{figure*}[h]
\centerline{\includegraphics[width=0.745\textwidth]{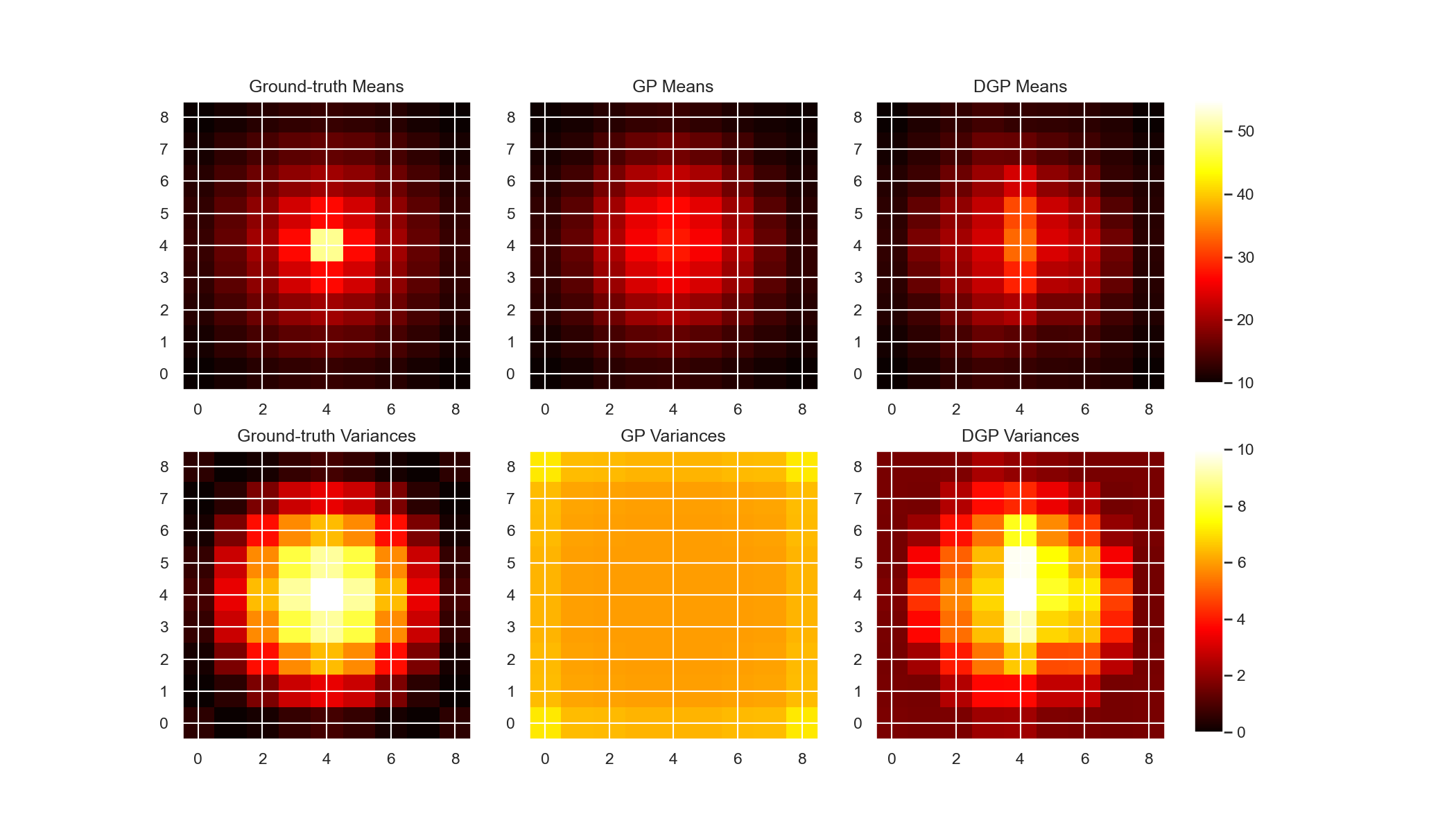}}
\caption{Radio Maps for the Central AP of the Chessboard Simulation}
\label{fig:cb_comparisons}
\end{figure*}

\section{Deep Gaussian Processes}\label{sec:dgp}

The restriction of GP in modelling heteroscedasticity is well studied and many fixes \cite{le2005heteroscedastic,kersting2007most,lazaro2011variational} have been proposed to address the issue. By stacking several GPs, DGP is also able to offer a heteroscedastic predictive uncertainty. 


While still largely in its infancy, DGP has already attracted attention from the signal fingerprinting community \cite{teng2018localization,wang2018deepmap,wang2020indoor}. In this work, we are paying attention to its ability in producing a more trustworthy predictive uncertainty. Without loss of generality, we here demonstrate how a DGP with two hidden layers can be utilised to achieve uncertainty modelling and argue that the major contribution of DGP sits in its capability of measuring the uncertainty of RSSI signals.
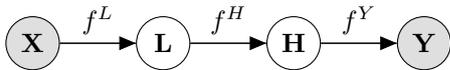
\begin{figure}[h]
\centerline{\begin{tikzpicture} \node[obs] (y) {$\mathbf{Y}$} ; %
        \node[latent, left=of y] (h) {$\mathbf{H}$} ; %
        \node[latent, left=of h] (l) {$\mathbf{L}$} ; %
        \node[obs, left=of l] (x) {$\mathbf{X}$} ; %
        \edge {x} {l} ; \edge {l} {h}; \edge {h} {y} ; 
        \draw (x)  --  (l) node [midway,above](TextNode){$f^L$};
        \draw (l)  --  (h) node [midway,above](TextNode){$f^H$};
        \draw (h)  --  (y) node [midway,above](TextNode){$f^Y$};\end{tikzpicture}}
\caption{Graphical model of a DGP with two hidden layers}
\label{fig:dgp}
\end{figure}

As is depicted in Fig~\ref{fig:dgp}, DGP corresponds to a graphical model with three kinds of nodes: the set of the positions of the surveyed fingerprints $\mathbf{X}\in \mathbb{R}^{N\times 2}$, the intermediate latent spaces $\mathbf{L}\in \mathbb{R}^{N\times D_L}$, $\mathbf{H}\in \mathbb{R}^{N\times D_H}$ and the corresponding RSSI observations $\mathbf{Y}\in \mathbb{R}^N$ for a single AP\footnote{For unsupervised DGP, $\mathbf{X}$ is an unobserved latent variable. In this paper, we only consider our application in signal fingerprinting, where $\mathbf{X}$ is the observed location tags.}. Compared to a single GP described in Equation \ref{eq:gp_noise}, for the n-th fingerprint $\mathbf{x}_n$, the generative process of DGP takes the form:
\begin{flalign}\label{eq:dgp_generate}
\begin{split}
y_n &= f^Y(\mathbf{h}_n) + \epsilon^Y_n \\
    h_{nd} &= f^H_d(\mathbf{l}_n) +\epsilon^H_n,\quad d=1,...,D_H\\
	l_{nd} &= f^L_d(\mathbf{x}_n) +\epsilon^L_n,\quad d=1,...,D_L
 \end{split}
\end{flalign}
where $f^L_d$s, $f^H_d$s and $f^Y$ are standalone GPs such that $f_d^L\sim \mathcal{GP}(\mathbf{0}, k^L_d(\mathbf{X}, \mathbf{X}))$ for $d=1,...,D_L$, $f_d^H\sim \mathcal{GP}(\mathbf{0}, k^H_d(\mathbf{X}, \mathbf{X}))$ for $d=1,...,D_H$ and $f^Y\sim \mathcal{GP}(\mathbf{0}, k^Y(\mathbf{L}, \mathbf{L}))$. The $f_d^L$s and $f_d^H$s are conditionally independent given $\mathbf{x}_n$. 

For DGP, the automatic relevance determination (ARD) covariance functions are commonly used:
\begin{equation}\label{eq:ard_kernel}
	k(\mathbf{x}_p, \mathbf{x}_q) =\sigma_{ard}^2\exp(-\frac{1}{2}\sum_{d=1}^Dw_d(x_{id}-x_{jd})^2)
\end{equation}
where $\sigma_{ard}$ plays the same role as $\sigma_f$ in Equation \ref{eq:gp_kernel}, and $w_d$ can be regarded as a dimension-wise length-scale. After optimisation, some of the $w_d$ might be very small, achieving the feature selection (structure selection) automatically.

Similar to GP, the training procedure of DGP requires optimization of the log likelihood:
\begin{equation}\label{eq:ll_dgp}
\log p(\mathbf{Y}|\mathbf{X})=\log \int_{\mathbf{L}, \mathbf{H}}p(\mathbf{Y}|\mathbf{H})p(\mathbf{H}|
\mathbf{L})p(\mathbf{L}|\mathbf{X})
\end{equation}
which is not analytically tractable because of the nonlinearity of $f^L$ and $f^Y$. To make it tractable, many methods such as variational inference \cite{damianou2013deep}, expectation propagation \cite{bui2016deep} and Monte-Carlo sampling \cite{vafa2016training} are applied. 

\begin{figure}[h]
\centerline{\begin{tikzpicture} \node[obs] (y) {$\mathbf{Y}$} ; %
        \node[latent, left=of y] (fy) {$\mathbf{F}^Y$} ; %
        \node[latent, left=of fy] (h) {$\mathbf{H}$} ; %
        \node[latent, left=of h] (fh) {$\mathbf{F}^H$} ; %
        \node[latent, left=of fh] (l) {$\mathbf{L}$} ; %
        \node[obs, above=of l] (x) {$\mathbf{X}$} ; %
        \node[latent, above=of fh] (uh) {$\mathbf{U}^H$} ; %
        \node[const, above=of uh] (tl) {$\tilde{\mathbf{L}}$} ; %
        \node[latent, above=of fy] (uy) {$\mathbf{U}^Y$} ; %
        \node[const, above=of uy] (th) {$\tilde{\mathbf{H}}$} ; %
        \edge {x} {l}  ; \edge {tl} {uh} ;\edge {uh} {fh} ; \edge {th} {uy} ;\edge {uy} {fy} ;\edge {l} {fh} ;\edge {fh} {h} ;\edge {h} {fy} ;\edge {fy} {y} ;  
        \end{tikzpicture}}
\caption{Graphical model representation of a DGP with inducing variables and two hidden layers}
\label{fig:dgpvi}
\end{figure}
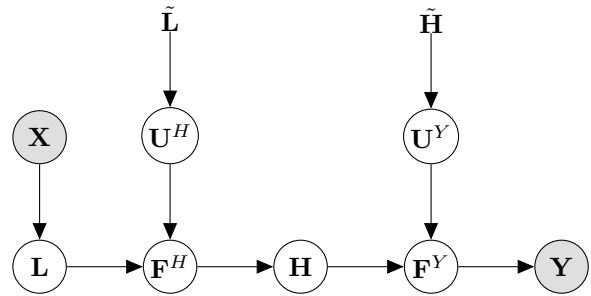

Following \cite{damianou2013deep} and \cite{titsias2010bayesian}, we apply sparse variational inference to optimise the DGP in this work. At first, we introduce $K_L + K_H$ inducing points $\tilde{\mathbf{L}}\in \mathbb{R}^{K_L\times D_L}$ and $\tilde{\mathbf{H}}\in \mathbb{R}^{K_H\times D_H}$ that correspond to their function values $\mathbf{U}^H\in\mathbb{R}^{K_L\times D_H}$ and $\mathbf{U}^Y\in\mathbb{R}^{K_H}$ respectively, which is illustrated in Fig~\ref{fig:dgpvi}. For clarity, we drop the conditioning on $\tilde{L}$ and $\tilde{H}$ for the remainder of this section.
Then we get the variational lower bound $\mathcal{F}_v\leq \log p(\mathbf{Y}|\mathbf{X})$ by Jensen's inequality:
\begin{equation}\label{eq:elbo}
\mathcal{F}_v=\int\mathcal{Q}\log\frac{p(\mathbf{Y}, \mathbf{F}^Y, \mathbf{H}, \mathbf{F}^H, \mathbf{L}, \mathbf{U}^H,\mathbf{U}^Y | \mathbf{X})}{\mathcal{Q}}
\end{equation}
where the integration is with respect to $\{\mathbf{L}, \mathbf{H}, \mathbf{F}^H \mathbf{F}^Y, \mathbf{U}^H, \mathbf{U}^Y\}$ and $\mathcal{Q}$ is the variational distribution such that
\begin{flalign}\label{eq:q}
\begin{split}
\mathcal{Q}=&p(\mathbf{F}^Y|\mathbf{U}^Y, \mathbf{H})q(\mathbf{U}^Y)q(\mathbf{H})\\
\cdot & p(\mathbf{F}^H|\mathbf{U}^H, \mathbf{L})q(\mathbf{U}^H)q(\mathbf{L})
 \end{split}
\end{flalign}
Here, $q(\mathbf{U}^Y)$ and $q(\mathbf{U}^H)$ are free-form variational distributions. $q(\mathbf{H})$ and $q(\mathbf{L})$ are Gaussian distributions factorized with respect to dimensions:
\begin{equation}\label{eq:qh}
q(\mathbf{H}) = \prod_{q=1}^{D_H}\mathcal{N}(\mathbf{\mu}_q^H, \mathbf{\sigma}_q^H),\quad q(\mathbf{L}) = \prod_{q=1}^{D_L}\mathcal{N}(\mathbf{\mu}_q^L, \mathbf{S}_q^L)
\end{equation}
where $\mathbf{S}_q^L$ is a full $N\times N$ matrix. In practice, to reduce the number of the variational parameters, $\mathbf{S}_q^H$ are commonly chosen to be diagonal. As $\mathbf{X}$ is given in this supervised scenario, we model $q(\mathbf{L})$'s covariance explicitly.

By expanding
\begin{flalign}\label{eq:expand}
\begin{split}
p(\mathbf{Y}, &\mathbf{F}^Y, \mathbf{H}, \mathbf{F}^H, \mathbf{L}, \mathbf{U}^H,\mathbf{U}^Y | \mathbf{X}) = \\
&p(\mathbf{Y}|\mathbf{F}^Y)p(\mathbf{F}^Y|\mathbf{U}^Y,\mathbf{H})p(\mathbf{H}|\mathbf{F}^H)p(\mathbf{F}^H, \mathbf{U}^H, \mathbf{L})p(\mathbf{L}|\mathbf{X})
 \end{split}
\end{flalign}
and substituting (\ref{eq:q}) and (\ref{eq:expand}) into (\ref{eq:elbo}), the variational lower bound $\mathcal{F}_v$ is of the form:
\begin{flalign}\label{eq:finalelbo}
\begin{split}
\mathcal{F}_v &= \left < \log p(\mathbf{Y|\mathbf{F}^Y}) + \log\frac{p(\mathbf{U}^Y)}{q(\mathbf{U}^Y)} \right >_{p(\mathbf{F}^Y|\mathbf{U}^Y,\mathbf{H})q(\mathbf{U}^Y)q(\mathbf{H})}\\
&+ \left < \log p(\mathbf{H|\mathbf{F}^H}) + \log\frac{p(\mathbf{U}^H)}{q(\mathbf{U}^H)} \right >_{p(\mathbf{F}^H|\mathbf{U}^H,\mathbf{L})q(\mathbf{U}^H)q(\mathbf{L})}\\
&+ \mathcal{H}_{q(\mathbf{H})} - \text{KL}(q(\mathbf{L}) || p(\mathbf{L} | \mathbf{X}))
 \end{split}
\end{flalign}
where $\left < \cdot \right >$ is the expectation, $\mathcal{H}$ denotes the entropy of a distribution and $\text{KL}(\cdot || \cdot)$ represents the Kullback - Leibler divergence. By optimising Equation \ref{eq:finalelbo}, we are able to obtain trained hyperparameters of the GPs introduced in Equation \ref{eq:dgp_generate}. 

The predictive variances by DGP are by nature heteroscedastic. The last layer (in fact every single layer) follows the basic GP as described in Section \ref{sec:gp} and the standalone GP here also produces the predictive variances that depend highly on the input coverage. However, by stacking GPs, the inputs of the last layer are the outputs of the hidden layer, whose data coverage has been shifted from the original input space (location tags) towards the output space (RSSI observations), resulting in a grouping or concentration in the hidden space if the RSSI observations are similar. Therefore, if the uncertainty is at some point low, the local coverage in the hidden space will be on the contrary high, and in the end, captured by the GP of the last layer as a low predictive variance. Intuitively, the stacked (deep) structure helps us to capture heteroscedasticity by propagating and representing the uncertainty in the hidden space.

\section{Evaluation}\label{sec:eval}

In this section, we show how GP and DGP differ in terms of uncertainty measurement via simulated fingerprint datasets where the underlying uncertainty is known. We further show how positioning performance can be affected by these two approaches with fingerprint datasets collected at a typical office building and a popular shopping mall. We follow what has been described in the previous sections in building radio maps and positioning fresh fingerprints. For specific implementation related suggestions, we recommend our readers to works like \cite{KUMAR20161,yiu2016locating,wang2018deepmap,wang2020indoor}.

\subsection{Simulated Chessboard}

Assume a chessboard with 9 rows and 9 columns, we simulate fingerprints at these 81 grids as if the signal observations were sampled from Gaussian distributions that are unique to every grid, where the mean and the variance on the grid $g$ are generated in the following way:
\begin{equation}\label{eq:cb_mean}
	\mu_{g} = \begin{cases}50 & \text{ if } ||X_g - X_{ap}||\leq 1 \\ 50 - 10log(10||X_g - X_{ap}||) & \text{ otherwise}\end{cases}
\end{equation}
\begin{equation}\label{eq:cb_Var}
    \sigma^2_{g} = 10cos^2(\frac{||X_g - X_{ap}||}{\pi})
\end{equation}

where $||X_g - X_{ap}||$ denotes the Euclidean distance between the AP and the grid. Here we are simulating a signal source that is isotropic and the signal attenuates faster when closer to the source (dictated by the logarithm term in Equation \ref{eq:cb_mean}). Especially, the signal is also heteroscedastic that it has higher uncertainty when closer to the source (dictated by the cosine term in Equation \ref{eq:cb_Var}). We placed 5 such simulated APs at the cebtre and four corners of the chessboard and draw samples from these APs to form fingerprints. At each grid, we collected 10 fingerprints for training and an extra fingerprint for testing. 

We build radio maps for these APs, using GP and DGP according to Section \ref{sec:gp} and Section \ref{sec:dgp} respectively, and locate the test fingerprints based on these radio maps. Notice that as these signals are artificially generated, we actually have the ground-truth radio maps for both the mean and the variance of all APs.

Take the central AP for an example, the ground-truth radio maps and the predictive maps generated by GP and DGP, for both means and variances can be found in Figure \ref{fig:cb_comparisons}. It is clear that the GP fails to capture the original uncertainties while DGP successfully recovers the fact the simulated signal has higher uncertainty at the center. Notice that the predictive variances by GP of the other four APs are quite similar to that of the central GP with rather flat predictive variances as expected, since they dominantly reflect the fingerprints coverage instead of the uncertainty as explained in Section \ref{sec:gp}. DGP captures about the right variances at the central area but predicts a slightly higher variance at the edges, probably due to the lack of more data beyond the chessboard area. 

\begin{figure}[h]
\centerline{\includegraphics[width=0.47\textwidth]{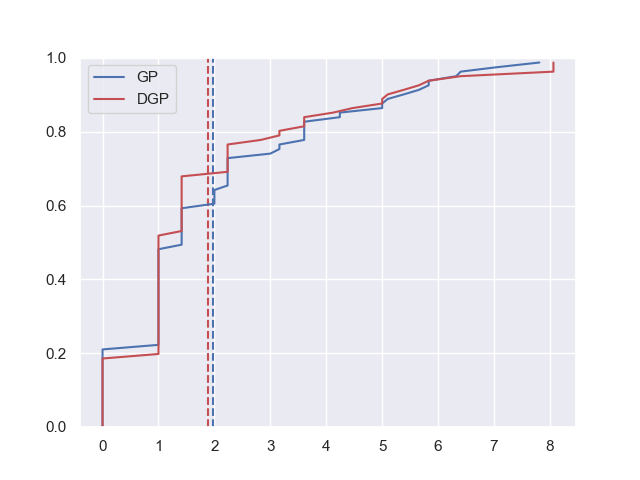}}
\caption{CDFs of the Chessboard Simulation}
\label{fig:cb_cdfs}
\end{figure}

As for the final positioning performance, DGP (mean accuracy 1.88) slightly outperforms GP (mean accuracy 1.97). The empirical cumulative distribution functions (CDF) of these two approaches can be found in Figure \ref{fig:cb_cdfs}. Since the uncertainty modelled by GP essentially stays almost the same at each grid, it degenerates Equation \ref{eq:likeli} to a term that mostly depends on the difference between the estimated mean and the observed signal measurement. However, this degeneration is unique in this setting where every grid is well surveyed but in more practical settings where the coverage is imbalanced across areas, the incorrect handling of variances will lead to bias towards surveyed areas. This also explains for the marginal differences between two approaches.

While the simulation can be tweaked in many ways, we would like to emphasise that the results of the chessboard simulation are qualitative only. Regardless of the exact toning of the simulation, GP will not work as intended when it comes to signal uncertainty measurement and the correct handling of the uncertainty by DGP improves positioning results. 

\subsection{V6 Office Building} 

\begin{figure}[h]
\centerline{\includegraphics[width=0.48\textwidth]{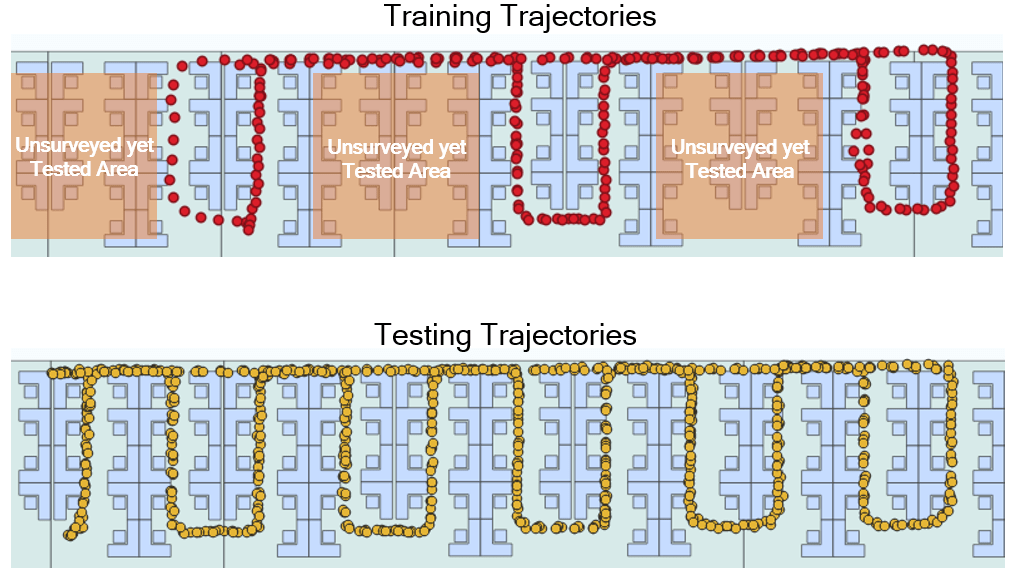}}
\caption{The V6 Office Building Dataset}
\label{fig:v6}
\end{figure}

\begin{figure*}[ht]
     \centering
     \begin{subfigure}[b]{0.32\textwidth}
         \centering
         \includegraphics[width=\textwidth]{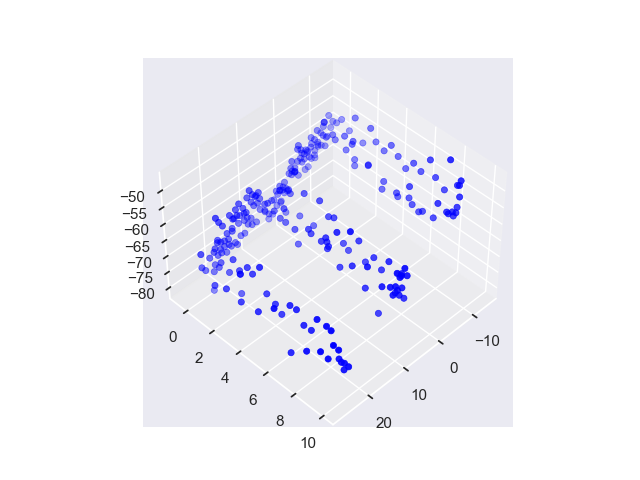}
         \caption{Raw Data Points}
         \label{fig:v6_raw_dp}
     \end{subfigure}
     \begin{subfigure}[b]{0.32\textwidth}
         \centering
         \includegraphics[width=\textwidth]{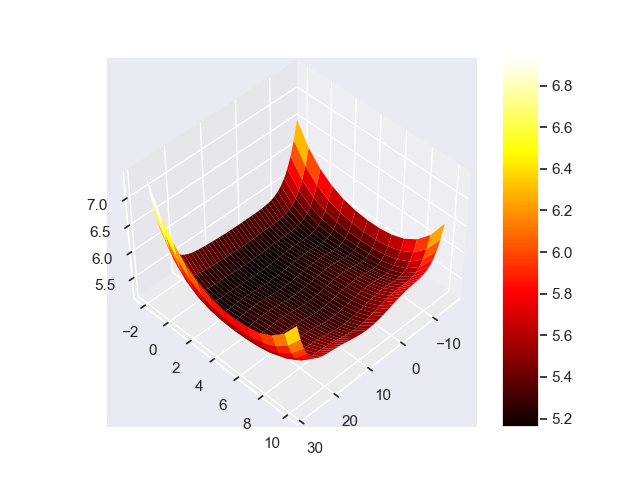}
         \caption{GP}
         \label{fig:v6_gp_var}
     \end{subfigure}
     \begin{subfigure}[b]{0.32\textwidth}
         \centering
         \includegraphics[width=\textwidth]{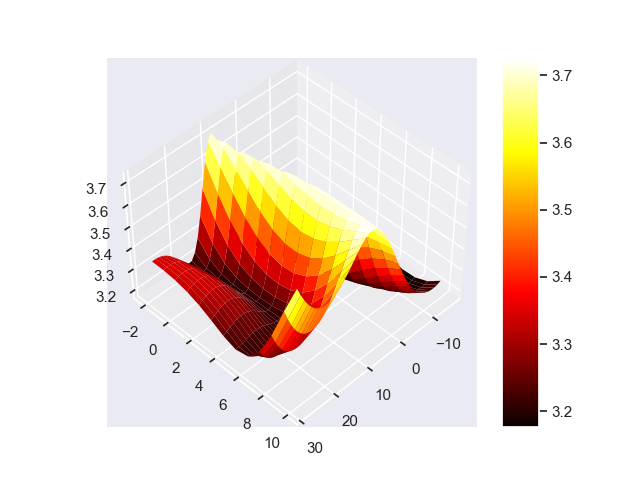}
         \caption{DGP}
         \label{fig:v6_dgp_var}
     \end{subfigure}
    \caption{Raw Data Points and Variances of Predictive Distributions from one AP of the V6 Dataset}
    \label{fig:v6_ap}
\end{figure*}

Practically, maintaining a good and balanced coverage is expensive and path survey \cite{gao2016easing} is a more attractive way to collect fingerprints. We path-surveyed one wing (covering an area with a size of \SI{45}{\meter} $\times$ \SI{14}{\meter}) of the first floor of a typical office building (V6) during working hours within a day, where there was a normal level of human activities during the survey. A Huawei Mate 30 Smartphone was used as the fingerprint collecting device while a proprietary vision based simultaneous localization and mapping device \cite{biswas2011corrective} similar to Google Tango was also carried by the surveyor as the ground-truth device for annotating the fingerprints with the correct location tags. Both devices are synchronised via the Network Time Protocol (NTP) service.

The training set consists of 2 trajectories while the test set consists of 4 trajectories. The training set does not cover all aisles between cubicles while the test set does. We believe this is especially effective for comparing the predictive power of the underlying approaches. Trajectories of the V6 dataset can be found in Figure \ref{fig:v6} and the vacuum areas are marked by shaded colour blocks in the training set.

We produced the radio maps (means and variances) using GP and DGP respectively. As expected, the variances produced by GP are shaped by the training trajectories. DGP models the uncertainty of the RSSI signal. An example of an AP located at the centre of the corridor is shown in Figure \ref{fig:v6_ap}. Variances by GP raise at the edges and the vacuum areas. Variances by DGP has a ridge at the centre, probably due to more human activities coming from the central cubicles. It is important to notice that the scales of these two radio maps are different. While the variances by GP do not seem to bound to any physical quantity, DGP models variances of slightly above 3, well matched with our observation for RSSI fluctuation (around 3 to 4 dB).

\begin{figure}[h]
\centerline{\includegraphics[width=0.47\textwidth]{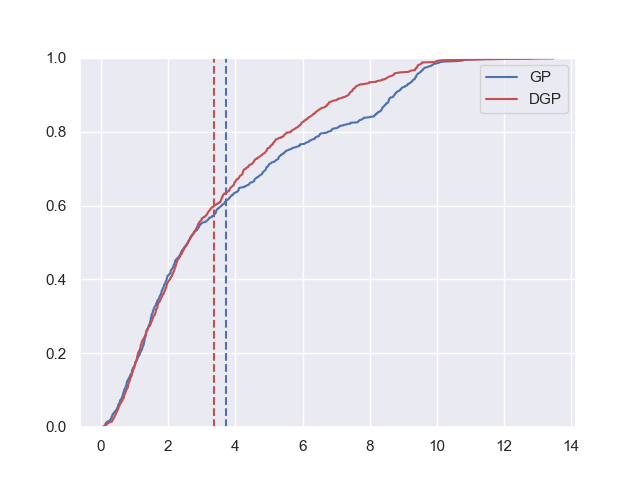}}
\caption{CDFs of the V6 Dataset}
\label{fig:v6_cdfs}
\end{figure}

The positioning CDFs can be found in Figure \ref{fig:v6_cdfs} where the dotted horizontal lines indicate the mean accuracy. GP produces a mean accuracy of \SI{3.71}{\meter} while DGP produces a mean accuracy of \SI{3.36}{\meter}. Although the improvement is marginal, it is observed that GP and DGP perform roughly the same at well surveyed areas and the vacuum area on the left-hand side (Figure \ref{fig:v6}), the major differences (from around the 55th percentile to 95th percentile in Figure \ref{fig:v6_cdfs}) come from the vacuum areas in the centre, where GP predicts higher variances due to lack of fingerprint coverage. This is effectively introducing a bias against the poorly surveyed area, which also means the claimed predictive ability of the GP is highly restricted if variances produced by GP is directly used. DGP, on the contrary, does a better job in modelling signal uncertainty and provides an estimate that is not biased towards the surveyed area, thus performs better in the vacuum areas.

\subsection{JCX Shopping Mall} 

\begin{figure}[h]
\centerline{\includegraphics[width=0.40\textwidth]{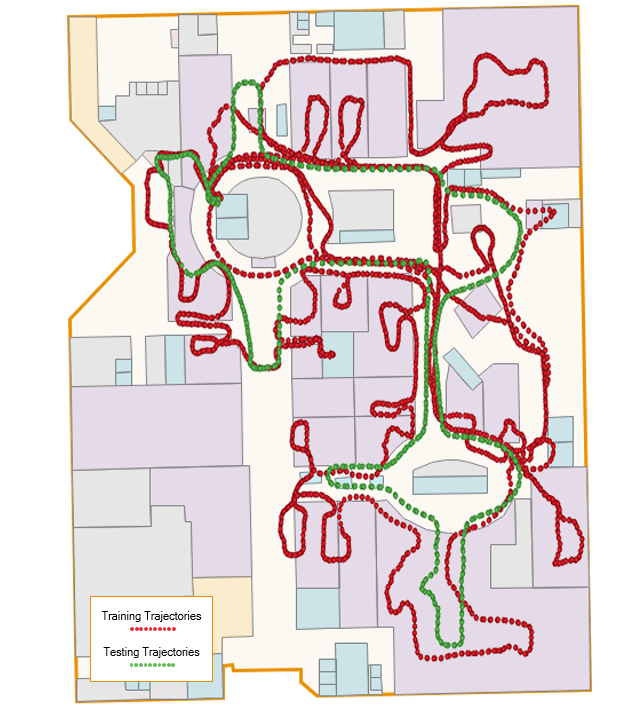}}
\caption{The JCX Shopping Mall Dataset}
\label{fig:jcx}

\centerline{\includegraphics[width=0.48\textwidth]{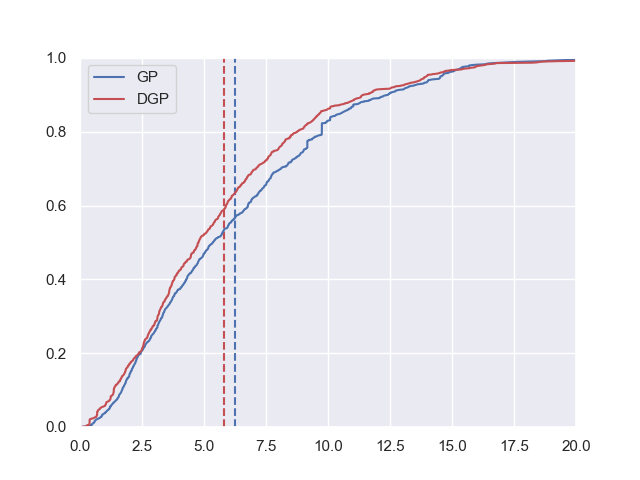}}
\caption{CDFs of the JCX Dataset}
\label{fig:jcx_cdfs}
\end{figure}

Similar to how we set up for the V6 datasets, we collected fingerprints from the ground floor (covering an area with a size of \SI{150}{\meter} $\times$ \SI{120}{\meter}) of a popular shopping mall (JCX) in Beijing during working hours within a day, where there were crowds shopping. We did not deliberately create vacuum areas in the training set this time and the trajectories roughly follow major crowds visiting routes and can be found in Figure \ref{fig:jcx}.

As for the positioning performance, the CDFs are plotted in Figure \ref{fig:jcx_cdfs}. GP produces a mean accuracy of \SI{5.79}{\meter} while DGP produces a mean accuracy of \SI{6.25}{\meter}. DGP slightly outperforms GP at all percentiles. Since testing trajectories are well covered by training fingerprints, we did not see a huge fingerprint coverage biased in the JCX dataset. Therefore the improvement comes largely from the correct modelling of uncertainty.


\section{Implications}\label{sec:conclusion}

We showed that the RSSI signal is heteroscedastic and the commonly practised GP does not model the uncertainty of the signal well, as the produced variances in fact link strongly to the coverage of the training fingerprints instead. As a result, the claimed advantages (predictive capabilities and uncertainty measurement) of GP are restrained. In contrast, DGP models uncertainty by broadcasting the uncertainty in latent layers and shows more trustworthy results.

Despite the fact that DGP theoretically avoids the pitfall of standalone GP in measuring uncertainty, it achieves so by trading off against computational power. In many applications, absolute accuracy may not be that important compared to other factors such as power consumption and reliability. Based on the above works, we conclude the following implications for designing a practical signal fingerprinting positioning system:

\begin{itemize}
    \item We showed a theoretical and practical advantage of DGP in measuring uncertainty, however, GP and DGP alike are in the end data-driven approaches. It is empirical to realise that the modelling of uncertainty needs training datasets of good quality and quantity. Additionally, the improvement of positioning accuracy could be marginal, given that the estimated means by GP and DGP are of the same accuracy. On the one hand, sparsely collected training fingerprints may not contain enough information for DGP to infer the uncertainty accurate enough to improve the positioning accuracy greatly. On the other hand, densely collected fingerprints pose a great challenge in terms of the training cost. Therefore, we recommend DGP over GP, only when the positioning accuracy is the highest design priority and the application allows trading efforts in collecting more fingerprints and computation cost for accurate uncertainty measurement. 

    \item For general applications, standalone GP is a decent option for building the radio maps of means. When the modelling of uncertainty is hard to be guaranteed due to lack of data, we recommend adopting uninformative variances rather than GP produced variances, and variance related GP computation (both during the radio map building and the positioning stages) can be skipped to save computational cost and avoid bias towards the surveyed area. If the positioning result needs to be constrained within the surveyed area, there are straightforward alternatives to involving the GP produced variances, such as simply reducing the grids of the chessboard to the surveyed area.
\end{itemize}

GP has been widely applied in signal fingerprinting yet our work demonstrates that GP gives incorrect uncertainty measurement. Measuring uncertainty from data is an active research topic in general and in the hope that more attention is given to uncertainty measurement within the signal fingerprinting research community, our preliminary work presents a rather simple two-layer DGP as a potential alternative. However, DGP has its own shortcomings such as high computational cost, our work can be further extended to adapt DGP into signal fingerprinting fully.

\bibliographystyle{unsrt}
\bibliography{gp2dgp}

\vspace{12pt}

\end{document}